\documentclass[lettersize,journal]{IEEEtran}
\usepackage{graphicx}
\usepackage{pifont}
\usepackage{algorithm}
\usepackage{mathtools}
\usepackage{algpseudocode}
\usepackage{pbox}
\usepackage{footmisc}
\usepackage{subcaption}
\usepackage{xr}
\usepackage{makecell}
\usepackage{tikz}
\usepackage{comment}
\usepackage{color,colortbl}
\usepackage{makecell}
\usepackage{booktabs}
\usepackage{float}
\usepackage{array}
\usepackage{setspace}
\usepackage{graphicx}
\usepackage{multirow,multicol}
\usepackage[]{amsmath}
\usepackage{amsbsy}
\usepackage{amssymb}
\usepackage{amsfonts}
\usepackage{pifont}
\usepackage{balance}
\usepackage{fancyvrb}

{\begin{list}               %    with flush left bullets.
    {$\bullet$ \hfill}{
        \setlength{\leftmargin}{\parindent}
        \setlength{\parsep}{0.04\baselineskip}
        \setlength{\itemsep}{0.5\parsep}
        \setlength{\labelwidth}{\leftmargin}
        \setlength{\labelsep}{0em}}
    }
{\end{list}}

\providecommand{\eref}[1]{\eqref{#1}}  % call \eqref from amstex
\providecommand{\cref}[1]{Chapter~\ref{#1}}

\providecommand{\fref}[1]{Figure~\ref{#1}}

\providecommand{\bydef}{\overset{\text{def}}{=}}

\renewcommand{\vec}[1]{\ensuremath{\boldsymbol{#1}}}

\providecommand{\calL}{\mathcal{L}}

\providecommand{\calP}{\mathcal{P}}

\providecommand{\calS}{\mathcal{S}}

\providecommand{\vh}{\mathbf{h}}

\providecommand{\vk}{\mathbf{k}}

\providecommand{\vv}{\mathbf{v}}

\providecommand{\vx}{\mathbf{x}}
\providecommand{\vy}{\mathbf{y}}

\newcommand{\cmark}{\ding{51}}
\newcommand{\xmark}{\ding{53}}

\providecommand{\vone}{\vec{1}}

\newcommand{\argmin}[1]{\mathop{\underset{#1}{\mbox{argmin}}}}

\newcommand\Tstrut{\rule{0pt}{2.6ex}}         % = `top' strut
\newcommand\Bstrut{\rule[-0.9ex]{0pt}{0pt}}  
\definecolor{Gray}{gray}{0.9}
\newcolumntype{g}{>{\columncolor{Gray}}c}

\begin{document}

\author{Yash~Sanghvi,~\IEEEmembership{Student~Member,~IEEE}, Abhiram~Gnanasambandam,~\IEEEmembership{Member,~IEEE},
Zhiyuan~Mao,~\IEEEmembership{Student~Member,~IEEE}, and~Stanley~H.~Chan,~\IEEEmembership{Senior~Member,~IEEE}%
\thanks{Y.~Sanghvi, Z.~Mao and S.~Chan are with the School of Electrical and Computer
Engineering, Purdue University, West Lafayette, IN 47907, USA. The work of A.~Gnanasambandam was completed when he was a graduate student at Purdue University.  Email: {
\{ysanghvi, mao114, stanchan\}}@purdue.edu,  abhiram.g94@gmail.com }
\thanks{The work is supported, in part, by the US National Science Foundation under the
grants IIS-2133032, and ECCS-2030570. }
}

\title{Photon-Limited Blind Deconvolution using Unsupervised Iterative Kernel Estimation}

\maketitle

\begin{abstract}
Blind deconvolution is a challenging problem, but in low-light it is even more difficult. Existing algorithms, both classical and deep-learning based, are not designed for this condition. When the photon shot noise is strong, conventional deconvolution methods fail because (1) the image does not have enough signal-to-noise ratio to perform the blur estimation; (2) While deep neural networks are powerful, many of them do not consider the forward process. When the noise is strong, these networks fail to simultaneously deblur and denoise; (3) While iterative schemes are known to be robust in the classical frameworks, they are seldom considered in deep neural networks because it requires a differentiable non-blind solver.

This paper addresses the above challenges by presenting an \emph{unsupervised} blind deconvolution method. At the core of this method is a reformulation of the general blind deconvolution framework from the conventional image-kernel alternating minimization to a purely kernel-based minimization. This kernel-based minimization leads to a new iterative scheme that backpropagates an unsupervised loss through a pre-trained non-blind solver to update the blur kernel. Experimental results show that the proposed framework achieves superior results than state-of-the-art blind deconvolution algorithms in low-light conditions.
\end{abstract}


\section{Introduction}
In low-light imaging applications such as microscopy \cite{pankajakshan2009blind,soulez2012blind,chen2013blind}, and astronomy \cite{jefferies1993restoration,schulz1993multiframe}, a consistent question being asked is how to estimate the blur kernel and deblur the image in the presence of photon shot noise. A natural formulation of the problem is the Poisson \emph{blind} deconvolution where the goal is to simultaneously recover the blur kernel $\vh$ and the latent image $\vx$ from the Poisson forward model
\begin{equation}
\vy = \text{Poisson}( \alpha \vh \circledast \vx ),
\end{equation}
where $\circledast$ denotes the convolution. The constant $\alpha$ here is a parameter that determines the mean photon level of the image $\vh \circledast \vx$. For low-light photography problems, $\alpha$ can be as low as a few photons per pixel, assuming that the latent image $\vx$ is normalized to the range of $[0,1]$.

\begin{figure*}[ht]
\includegraphics[trim={40 0 10 110},clip,width=\linewidth]{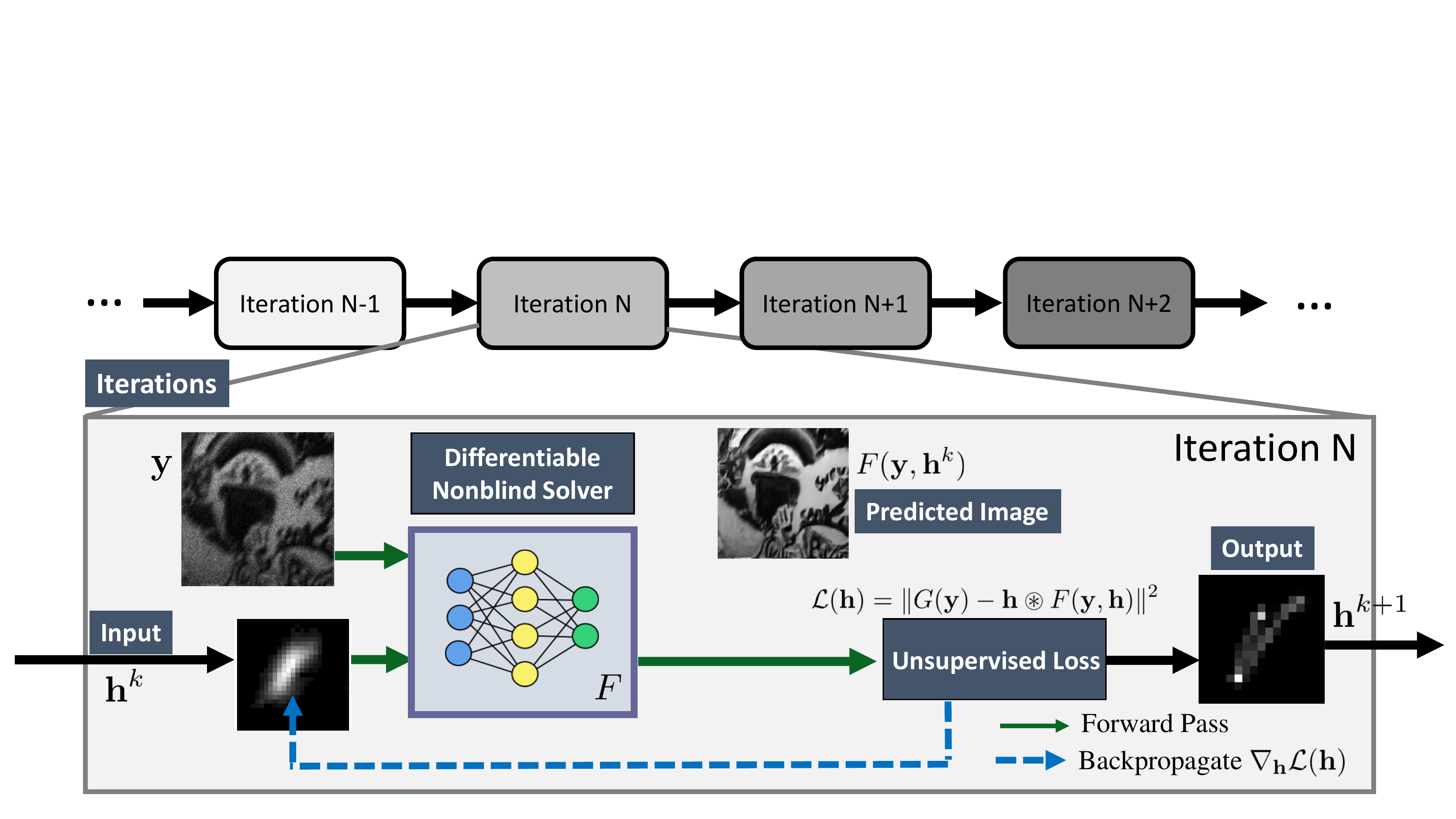}
\caption{\textbf{Proposed blind deconvolution framework.} The proposed method is unsupervised and iterative. At the $k$th iteration, the algorithm takes the current estimate of the blur kernel $\vh^k$ and feeds forward through a nonblind solver to obtain an image estimate $F(\vy,\vh^k)$. An unsupervised loss is then used to backpropagate and update the kernel. Since the nonblind solver is differentiable, the gradient of the loss can be computed.}
\end{figure*}

Because of the low-light condition (i.e., a small $\alpha$), a brute-force implementation of existing blind deconvolution algorithms, including fine-tuning state-of-the-art deep-learning models using the Poisson data, would face three challenges:

\begin{enumerate}
\item[(i)] \textit{Heavy noise makes kernel estimation hard}. Even in the noiseless case, the joint optimization of the blur kernel and the latent image is known to cause degenerate solutions according to Levin et al. \cite{levin2011understanding}. This finding is consistent with many previous algorithms that propose to find the blur kernel. For example, many propose to obtain the pilot estimates of the images, such as applying shock filters and extracting edges, and using these pilot estimates to update the kernel \cite{xu2010two,cho2009fast,shan2008high}. However, when the signal-to-noise ratio is low, the kernel estimation would be very challenging because no pilot estimates of sufficient quality can be constructed.
\item[(ii)] \textit{Blur agnostic deep neural networks seldom utilize the forward model to guide the restoration}. Generic deep-learning methods that are agnostic to the actual forward equation are known to perform worse than those explicitly taking into account the blurring process, as evident in learning-based linear inverse solvers \cite{monga2021algorithm}. This gap is further widened in heavy-noise conditions. For Poisson distributions, there are currently very few deep-learning methods that explicitly model them in the inversion process \cite{sanghvi2021photon,mur2021deep}.
\item[(iii)] \textit{Iterative methods are important but not used}. Many classical methods estimate the kernel iteratively because the alternating minimization strategy is known to be effective. Even for non-blind deconvolution, people found iterative methods perform well \cite{gong2020_LearningDeep,nan2020_VariationalEM}. Delbracio et al. \cite{delbracio2021polyblur} argued that iteratively deblurring the image leads to better and more stable solutions. However, this concept is missing in today’s deep-learning based blind deconvolution method because there is no fast, effective, and \emph{differentiable} non-blind solvers.
\end{enumerate}

\begin{figure}[h]
\centering
\begin{tabular}{ccc}
    \includegraphics[width=0.32\linewidth]{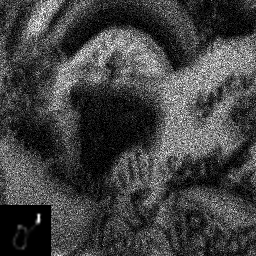} &
    \hspace{-2.0ex}\includegraphics[width=0.32\linewidth]{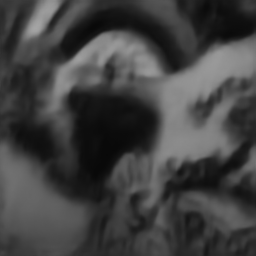} &
    \hspace{-2.0ex}\includegraphics[width=0.32\linewidth]{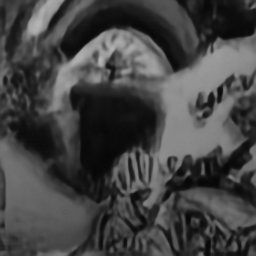} \\
    Blurred noisy image & MPR-Net \cite{zamir2021multi} & Proposed\\

\end{tabular}
\caption{ \textbf{Demonstration of the proposed method}. A Blurred image corrupted with Poisson noise (photon level $\alpha = 20$) is provided on the left. The blur kernel, unknown to the reconstruction methods, is provided in the inset. We take a state-of-the-art deblurring supervised method, i.e., MPR-Net \cite{zamir2021multi}, retrain it using Poisson blurred images, and show the corresponding reconstructions using MPR-Net (middle) and technique proposed in this paper (right).}
\end{figure}

Overcoming the challenges mentioned above requires a new design of the blind deconvolution algorithm. The main idea presented in this paper is a differentiable framework that back-propagates the loss to update the \emph{blur kernel} (instead of the model parameters), as illustrated in Figure 1. The framework is unsupervised because the non-blind solver shown in the middle of the figure is pre-trained and fixed. Once the non-blind solver is plugged in, there is no training. During testing, the blur kernel is updated by back-propagating an unsupervised loss which is computed by propagating the gradient of the non-blind solver. The blur kernel is updated iteratively instead of estimated directly by a separate module.

The proposed method has two advantages compared to the existing state-of-the-art deep neural network blind deconvolution algorithms: (1) It is unsupervised. This helps the algorithm to be adaptive to different blur kernels. (2) It is flexible as any differentiable non-blind solver can be plugged into this framework.

A snapshot of the main results of the proposed method is shown in Figure 2. In this example, the photon level is set to be as low as $\alpha = 20$ (equivalent to Gaussian noise of standard deviation $\sigma \approx 60/255$ in terms of the signal-to-noise ratio). This is a substantially noisier problem compared to the ones reported in the deconvolution literature. Compared to state-of-the-art methods such as the MPR-Net \cite{zamir2021multi} that has been fine-tuned using the Poisson data, the performance of the proposed method is better both in terms of the peak signal-to-noise ratio (PSNR) and the visual quality.

\section{Related Work}

\textbf{Non-blind deconvolution}. Non-blind deconvolution algorithms assume a known blur kernel \cite{banham1997digital}. For the Poisson case, methods have been studied in as early as the 1980s. For example, the Richardson-Lucy algorithm \cite{richardson1972bayesian,lucy1974iterative} is iterative method that converges to the maximum likelihood estimate (MLE) of the Poisson deconvolution problem. A variety of Maximum A Posteriori (MAP) based solutions with different regularization methods have been used to solve the problem \cite{figueiredo2010restoration,harmany2011spiral,nowak2000statistical}. PURE-LET \cite{purelet} estimates the clean image by minimizing the Poisson unbiased risk estimate. Iterative solutions using the ADMM scheme have also been proposed to incorporate Total Variation (TV), and complex image priors \cite{figueiredo2009deconvolution,figueiredo2010restoration,rond2016poisson}.

\textbf{Traditional blind deconvolution}. Single image blind deconvolution has a long list of work since the 1990s \cite{chan1998total}. A common theme is to jointly optimize for both the kernel and the image iteratively \cite{shan2008high,cho2009fast,xu2010two,mao2020}. Under this joint optimization framework, many image priors and kernel priors are utilized, such as total variation (TV) \cite{chan1998total}, local gradient maxima \cite{chen2019blind}, internal patch recurrence \cite{michaeli2014blind}, and sparse wavelet approximation \cite{cai2009blind}. Among those, a mixture of  $\ell_1$-norm and Total Variation (TV) have been proposed in \cite{Anger2019} to deblur images under a high noise regime.

As far as kernel estimation is concerned, people observe that it is more robust to use image gradients instead of the image itself \cite{shan2008high,joshi2008psf,cho2009fast,xu2013unnatural}. Methods such as \cite{xu2010two,gong2016blind} improve the joint optimization scheme by identifying the salient gradients of the image and using them in the kernel estimation process. These methods are often employed in a multi-scale manner, starting from a coarse kernel and image estimate, and using these estimates at a finer scale \cite{fergus2006removing,cho2009fast}.

Despite the empirical success, there is an issue in the framework. Levin et al. \cite{levin2009understanding,levin2011understanding} showed that the joint minimization can theoretically converge to a degenerate solution pair, i.e., it will recover an identity kernel instead of the true kernel. Levin et al.'s argument was that the joint minimization needs to simultaneously estimate the kernel and the image. The search space is too large for the available amount of observations. Their suggestion was to exploit the asymmetry of the problem by estimating the kernel first because the kernel has fewer variables than the image. So, it is easier to recover the kernel than jointly recovering the kernel and the image.

\textbf{Learning based blind deconvolution}. Since the last decade, deep learning methods have been used to solve the blind deconvolution problem. Different CNN-architectures such as Scale Recurrent Network (SRN) \cite{tao2018scale}, Deep-Deblur \cite{Nah_2017_CVPR}, MPR-Net \cite{zamir2021multi} have achieved impressive performance on single image deblurring without considering the forward model. Some learning based methods such as \cite{chakrabarti2016neural,gong2017motion,schuler2015learning,wang2019image} do the opposite by taking the forward imaging model into consideration. For example, \cite{chakrabarti2016neural} learns the Fourier coefficients of the deconvolution filter from patches of image and deconvolves the blurred image by applying the patchwise average of the predicted filter. \cite{sun2015learning} predicts the non-uniform motion blur field using a CNN and Markov Random Field (MRF) model. \cite{Agarwal2020} designs its network structures by unfolding the classic Richardson-Lucy algorithm. Another line of work is to use deep neural network as implicit prior for images or kernels \cite{Ren2020}, \cite{Fang2022}, which also achieves competitive performance in blind or non-blind image deblurring tasks.

\section{Iterative Kernel Estimation}
\subsection{Main Idea}
To explain the proposed approach, it would be useful to start with the classical blind deblurring algorithm for Gaussian noise. In the classical setting, the joint optimization of the blur kernel $\vh$ and the image $\vx$ is
\begin{equation}
    (\widehat{\vx},\widehat{\vh}) = \argmin{\vx,\vh} \; \bigg\{\|\vy - \vh \circledast \vx\|_2^2 + \lambda R(\vx) + \gamma S(\vh)\bigg\},
\end{equation}
for some regularization functions $R(\vx)$ and $S(\vh)$, and $\circledast$ denotes the convolution operator. The standard strategy to solve the joint optimization is to alternate between $\vx$ and $\vh$ by fixing one and updating the other.

For Poisson noise, the likelihood function changes from the $\ell_2$-norm squares to the Poisson-likelihood:
\begin{equation}
\calP(\vy, \vh \circledast \vx) \bydef \vone^T(\vh \circledast \vx) - \vy^T\log(\vh \circledast \vx) + \vone^T \log \vy!
\end{equation}
where the last term $\vone^T \log \vy!$ can be dropped because it does not depend on $\vx$ and $\vh$. Following the same alternating minimization principle, the algorithm for the Poisson blind deblurring consists of two steps:
\begin{align}
\vx^{k+1} &= \underset{\text{non-blind Poisson deblurring}}{\underbrace{\argmin{\vx} \;\; \calP(\vy, \vh^k \circledast \vx) + \lambda R(\vx)}} \bydef F(\vy,\vh^k),
\label{eq: x-subproblem}\\
\vh^{k+1} &= \argmin{\vh} \;\; \calP(\vy, \vh \circledast \vx^{k+1}) + \gamma S(\vh). \label{eq: h-subproblem}
\end{align}
A key observation here is that the $\vx$-subproblem in \eref{eq: x-subproblem} is a Poisson proximal map using some regularization $R(\vx)$. We define this mapping as a function $F(\vy,\vh^k)$ which takes the corrupted image $\vy$ and the current estimate of the kernel $\vh^k$ to produce a deblurred image $\vx^{k+1}$. Since $\vh$ is known and fixed at $\vh^k$, the mapping $F(\vy,\vh^k)$ is a \emph{non-blind} Poisson deblurring method. As will be elaborated later, this non-blind Poisson deblurring step is implemented via a deep neural network.

The first step of the proposed method is to substitute $F(\vy,\vh^k)$ into the $\vh$-subproblem in \eref{eq: h-subproblem} and merge the pair of alternating equations \eref{eq: x-subproblem} and \eref{eq: h-subproblem} into one. This will give us
\begin{equation}
\vh^{k+1} = \argmin{\vh} \;\; \bigg\{\calP(\vy, \vh \circledast F(\vy,\vh^k
)) + \gamma S(\vh)\bigg\}.
\label{eq: step 1}
\end{equation}
Notice here \eref{eq: step 1} has no difference with the equation pair \eref{eq: x-subproblem} and \eref{eq: h-subproblem} as far as implementation is concerned: we fix $\vh^k$ and compute $F(\vy,\vh^k)$, and then we estimate the kernel by assuming a fixed $F(\vy,\vh^k)$.

\begin{figure*}[h]
\centering
\vspace{-2ex}
\includegraphics[width=\linewidth]{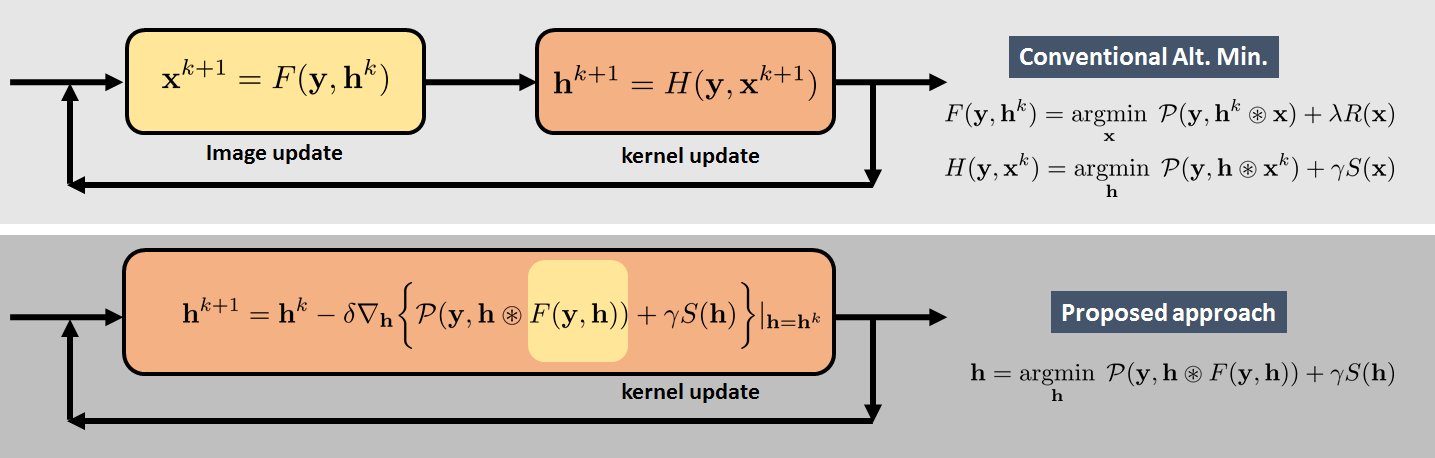}
\caption{\textbf{Conceptual illustration of the proposed method}. In conventional alternating minimization, we alternatingly update the image and the blur kernel through two separate modules. In the proposed approach, we leverage the differentiability of the non-blind Poisson solver and directly estimate the blur kernel.}
\label{fig: compare}
\vspace{-2ex}
\end{figure*}

The second and a very important step is to recognize that \eref{eq: step 1} is still a two-variable optimization because $F(\vy,\vh^k)$ is the non-blind solver: when we run $F(\vy,\vh^k)$, we are solving the optimization \eref{eq: x-subproblem}. The innovation here is to turn this two-variable alternating optimization in $(\vx,\vh)$ into a one-variable optimization in $\vh$. To do so, we recognize \eref{eq: step 1}  takes the form of a fixed-point equation --- given $\vh^k$, plug it into an equation, obtain $\vh^{k+1}$, and repeat. We propose to consider the \emph{equilibrium} of this fixed-point iteration. The equilibrium is obtained by dropping the indices $k$ and $k+1$ from the equation. This will lead to an optimization that does not involve the iteration index $k$:
\begin{equation}
\vh = \argmin{\vh} \;\; \bigg\{\calP(\vy, \vh \circledast F(\vy,\vh)) + \gamma S(\vh)\bigg\}.
\label{eq: step 2 equilibrium}
\end{equation}
Although in terms of notations, the difference between \eref{eq: step 2 equilibrium} and \eref{eq: step 1} is subtle, the physical significance is huge. \fref{fig: compare} shows a pictorial illustration. In \eref{eq: step 2 equilibrium}, the optimization is completely in $\vh$. The variable $\vx$ is never involved. The way to visualize is that while $F$ theoretically takes the form of an optimization which is \eref{eq: x-subproblem}, it is nonetheless just a neural network. As long as we know the input-output relationship of this neural network, it can be absorbed as a part of the forward image formation model. Because it is now part of the image formation model, the optimization in $\vh$ just needs to know how to take gradients of this nonlinear forward model.

At this point, one may anticipate that a reasonable algorithm would be to run a gradient descent step to minimize the objective function. However, there is a caveat of adopting a gradient-based algorithm --- the Poisson-likelihood is not differentiable at zero and hence it will cause difficulties. Besides, \eref{eq: step 2 equilibrium} asks us to estimate the kernel $\vh$ directly from the noisy image $\vy$. This is not a good approach because the Poisson noise is often very strong relative to the signal. Even in the classical literature, people have proposed to estimate $\vh$ from some variants of $\vy$ (such as pixels around the strong edges) instead of the original noisy-blurry image $\vy$.

Following the above argument, the proposed method introduces the third step by applying a Poisson \emph{denoiser} $G$ to the input noisy image. That is, we define
\begin{equation}
    G(\vy) = \argmin{\vx} \;\; \bigg\{ \vone^T\vx - \vy^T\log(\vx) + \lambda R(\vx) \bigg\},
    \label{eq: G}
\end{equation}
for some regularization function $R(\vx)$. Just like the non-blind deblurring step, $G(\vy)$ can be implemented via a neural network.

The next big observation here is that while $\vy$ is the Poisson count that ranges from zero to infinity, the estimate $\vx$ is normalized to the range of $[0,1]$. Therefore, while $\vy$ exhibits the Poisson characteristics, the estimate $G(\vy)$ does not have any Poissonian structure and only contains the blur. The ``noise'' remaining in $G(\vy)$ is algorithm-dependent because a different denoising algorithm will generate a different residue pattern. From a statistical point of view, we can empirically find out the distribution of the residue but practically this would require access to the ground-truth. However, since $G(\vy)$ is distributed in the range of $[0,1]$, it would match with the range of the prediction $\vh \circledast F(\vy,\vh)$. This gives us an approximated objective function:
\begin{equation}
\vh = \argmin{\vh} \;\; \bigg\{\| G(\vy) - \vh \circledast F(\vy,\vh))\|_2^2 + \gamma S(\vh)\bigg\}.
\label{eq: step 3 estimate h}
\end{equation}
The choice of the $\ell_2$-norm squares is more of convenience than a rigorous statistical reasoning because we do not know the distribution of the residue $G(\vy) - \vh \circledast F(\vy,\vh)$. However, this slackness does not seem to cause significant drawbacks in our experimental results.

The optimization in \eref{eq: step 3 estimate h} is the core of the proposed idea. Comparing this with the alternating minimization in \eref{eq: x-subproblem} and \eref{eq: h-subproblem}, the new formulation has two advantages:

\begin{enumerate}
\item \textbf{Optimization in Kernel Space}. The new formulation pushes all the efforts to estimating the kernel. This is consistent with the literature such as \cite{cho2009fast,xu2010two} that also spend most of the time trying to obtain a good kernel. Also, as suggested in Levin et al. \cite{levin2011understanding}, we perform estimation in the smaller kernel space $\vh$ instead of joint estimation which has shown to converge to the no-blur degenerate solution.
\item \textbf{Unsupervised Loss} \eref{eq: step 3 estimate h} enables \emph{unsupervised} estimation of $\vh$. Note that the neural network $F$ and $G$ and pre-trained and fixed under the proposed framework. No ground truth is needed as far as the estimation of the kernel is concerned.
\end{enumerate}

\subsection{Choice of $F$ and $G$}
Since the premise of the problem is low-light, the Poissonian structure has to be handled. In the proposed approach, the Poisson part is realized through the choice of $F$ and $G$.

For the choice of $F$, its goal is to take a blur kernel $\vh$ and a noisy-blurry image $\vy$ to return an deblurred image $F(\vy,\vh)$, which is essentially \eref{eq: x-subproblem}. There are many non-blind Poisson deconvolution solvers. This paper uses a recent method by Sanghvi et al. \cite{sanghvi2021photon}. In this approach, the minimization is solved via an algorithm unrolling that unrolls the plug-and-play alternating direction method of multiplier (PnP-ADMM) into a chain of repeated blocks of steps. The advantage of the unfolded network is that $F(\cdot)$ constructed in this way will be \emph{differentiable}, a key property required to solve \eref{eq: step 3 estimate h}.

\begin{figure}
    \centering
    \includegraphics[width=0.99\linewidth]{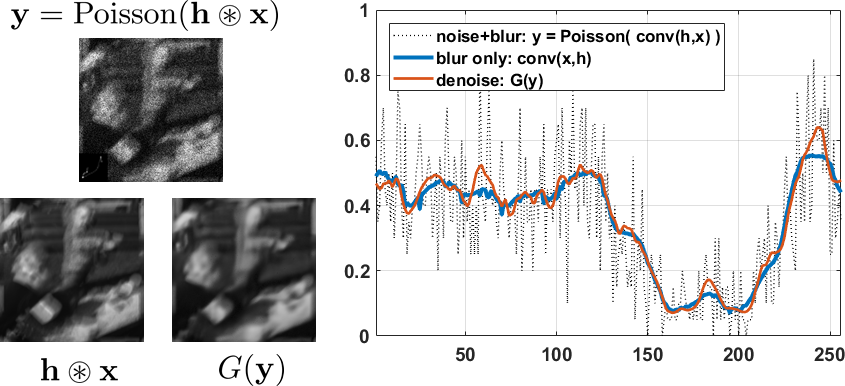}
    \caption{\textbf{Why use $G(\vy)$?} (Left) Using the denoiser $G(\vy)$, we aim to remove the Poisson noise from the blurred image. This image serves as a target for the data fidelity term in the cost function we optimize later. (right) 1D scan of the middle row of the 3 images provided on the left. $G(\vy)$ removes the noise from the image $\vy$ but preserves the blurred edges.}
    \label{fig:denoiser_output}
\end{figure}

The denoiser $G$ follows equation \eref{eq: G}. $G$ is a special case of $F$ where there is no blur kernel. Unlike a generic denoiser, $G$ is specifically trained to remove the Poisson noise from the noisy-blurred input and keep the blur intact so it can serve as a target for the estimated blurred image i.e. $\vh \circledast F(\vy, \vh)$. The impact of $G$ is illustrated in \fref{fig:denoiser_output}. We consider a noisy-blurry image $\vy = \text{Poisson}(\alpha \vx)$ and take a cross-section of the pixels. Because of the noise, the cross-section plot is extremely noisy and blurry. The Poisson denoiser $G$ will remove the noise but the blur along the edges is preserved. As a result, estimating the blur kernel from $G(\vy)$ would become easier.

The regularization function $\lambda R(\vx)$ used to define $F$ and $G$ is a symbolic place-holder. During the actual implementation, since both $F$ and $G$ are the unfolded PnP-ADMM, the regularization function $\lambda R(\vx)$ is realized via a denoising network, as documented in several prior work \cite{venkatakrishnan2013plug,sreehari2016plug,ahmad2020plug}. The exact value of the parameter $\lambda$ and the explicit form of the regularization $R(\vx)$ are unimportant.

\subsection{Iterative Scheme}
We now discuss how to solve the minimization in \eref{eq: step 3 estimate h}. Define the following as the data fidelity term:
\begin{equation}
    \calL(\vh) \bydef \| G(\vy) - \vh \circledast F(\vy,\vh)) \|_2^2
    \label{eq:cost_function}
\end{equation}
so that the overall optimization becomes
\begin{equation*}
\vh = \argmin{\vh} \;  \underset{\text{Overall Loss}}{\underbrace{\calL(\vh) + \gamma S(\vh)}},
\end{equation*}
where we define the overall loss as the sum of the data fidelity term $\calL(\vh)$ and the regularization $\gamma S(\vh)$.

To decouple the minimization of the first term $\mathcal{L}(\vh)$ from the minimization of $\mathcal{S}(\vh)$ (which can potentially be non-differentiable such as $\calS(\vh) = \|\vh\|_1$), we perform variable splitting as done in a standard manner in half-quadratic splitting. More specifically, we introduce an additional variable $\vv$ and enforce the constraint $\vh =\vv$ using a quadratic penalty. This leads us to the following iterative steps to minimize the cost function in \eqref{eq:cost_function}
\begin{align}
    \vh^{k+1} &= \argmin{\vh} \; \calL(\vh) + \frac{\mu}{2}\|\vh - \vv^k\|_2^2, \label{eq update h}\\
    \vv^{k+1} &= \argmin{\vv} \;\; \gamma S(\vv) + \frac{\mu}{2}\|\vv - \vh^{k+1}\|_2^2 \label{eq update v},
\end{align}
for some hyperparameter $\mu$.

The $\vh$-subproblem \eqref{eq update h} is solved using gradient descent. To save computation, $\vh$ can be updated \emph{inexactly} via one gradient descent step:
\begin{equation}
    \vh^{k+1} = \vh^k - \delta \cdot \Bigg\{ \underset{\text{backpropagate } F}{\underbrace{\nabla_{\vh}\big\{L(\vh)\big\}\Big|_{\vh = \vh^k}}} + \mu(\vh^k - \vv^k)\Bigg\},
\end{equation}
where $\delta$ is the gradient descent step size.
The gradient can be found using automatic differentiation, i.e., backpropagation through the non-blind deblurring algorithm. In popular deep learning packages such as \verb|PyTorch| and \verb|TensorFlow|, the implementation is done by the \verb|autograd| function in these packages. In one line, namely \verb|loss.backward()|, one can compute the gradients of all the variables involved in the calculation of the variable \verb|loss|.

For the $\vv$-subproblem \eqref{eq update v}, a natural choice of the regularization function is $S(\vv) = \|\vv\|_1$. Other regularization functions can also be used; however, if the problem of interest mainly concerns about the support of kernel such as motion blur, then $\ell_1$-norm is a reasonable choice. Computationally, if $S(\vv) = \|\vv\|_1$, there exists a closed-form solution using the shrinkage formula:
\begin{align}
    \vv^{k+1}
    &= \max\left(\left|\vh^{k+1}\right|-\gamma/\mu,0\right)\cdot \text{sign}(\vh^{k+1}) \notag \\
    &\bydef \calS_{\gamma/\mu}(\vh^{k+1}).
\end{align}

\subsection{Overall Algorithm and Initialization}
The overall algorithm is shown in Algorithm~\ref{alg:poiss_deconv}. The algorithm contains a few additional modifications to make the algorithm more robust. For example, the parameter $\mu$ and $\gamma$ are updated according to the following heuristics. $\mu$ is increased by a constant factor to ensure that the two variables representing the kernel estimate $\vh, \vv$ converge to the same value. $\gamma$, which represents the strength of the $\ell_1$ norm prior, is gradually decreased over the iterations to allow for sparse solutions initially and then for less sparse solutions in later iterations.  In our implementation, the operators $G$ and $F$ also requires a photon level estimate (equivalent to the noise level in the Gaussian case.)

Due to the ill-posedness of the blind deconvolution problem, the initialization of the kernel estimate plays an important role. The goal of our initialization schemes is to obtain a reasonable starting point which is computationally inexpensive. For this paper, we use the method provided in \cite{delbracio2021polyblur}. Specifically, the method assumes the blur kernel as a tilted anisotopic Gaussian kernel parametrized by 3 parameters - the major axis, minor axis, and tilt of the kernel. We use the denoised image $G(\mathbf{y})$ as input to the scheme and use the resulting anisotropic Gaussian kernel as initial kernel estimate $\vh^{0}$.

\begin{algorithm}[H]
\begin{algorithmic}[1]
\State \textbf{Input}: Noisy-blurry $\vy$, denoiser $G(\cdot)$, and non-blind solver $F(\cdot)$.
\State Denoise $\vy$ to obtain a noiseless blurred image $G(\vy)$
\State Initialize $\vh^0$ and $\vv^0$.
\State $\mu \leftarrow 2.0$, $\gamma \leftarrow 10^{-3}$
    \For{$k = 0, 1, 2, \cdot\cdot\cdot$}
    \State $\mathcal{L}(\vh) \leftarrow \|G(\vy) - \vh \circledast F(\vy,\vh)\|_2^2$
    \State Calculate $\nabla_{\vh}\mathcal{L}(\vh^k)$ using automatic differentiation
    \State $\vh^{k+1} \leftarrow \vh^{k} - \delta \big( \nabla_{\vh}\mathcal{L}(\vh^k) + \mu(\vh^k - \vv^k)  \big)$
    \State $\vv^{k+1} \leftarrow \mathcal{S}_{\gamma/\mu}(\vh^{k+1})$
    \State $\mu \leftarrow 1.01 \mu$, $\gamma \leftarrow \gamma/1.01$
    \EndFor
\State return $\vh^{(\infty)}$ and $\vx^{(\infty)} = F(\vy,\vh^{(\infty)})$
\end{algorithmic}
\caption{Iterative Poisson Deconvolution Scheme}
\label{alg:poiss_deconv}
\end{algorithm}

\section{Experiments}
\begin{table*}[]
    \setlength\doublerulesep{0.5pt}
    \centering
    \begin{tabular}{p{6em}gccccggg|c}
        \toprule
        \multirow{2}{*}[0.5em]{Method $\rightarrow$} &  Two-Phase  &  SRN &  DMPHN & Deep-Deblur  & MPRNet & Poisson PnP & PURE-LET & Ours & P4IP  \Tstrut\\
        Photon lvl $\downarrow$ & \cite{xu2010two} & \cite{tao2018scale} &
        \cite{zhang2019deep} &
        \cite{Nah_2017_CVPR} &  \cite{zamir2021multi} & \cite{rond2016poisson} &  \cite{purelet} & &\cite{sanghvi2021photon} \Bstrut\\
        \hline
        \rule{0pt}{2ex}
        \multirow{2}{*}[0.2em]{$\alpha = 10$} & 16.42 & 20.33 & 20.25 & 20.92  & 21.03  & 19.83 & 21.63 & \textbf{22.10} & 22.45  \\
        \vspace{0.2ex}
         & 0.511 & 0.511 & 0.509 & 0.523 & 0.533  & 0.464 & 0.607 &
         \textbf{0.598} & 0.639 \\
        \hline
        \rule{0pt}{2ex}
        \multirow{2}{*}[0.2em]{$\alpha = 20$} & 17.40 & 20.46 & 20.43 & 21.11 & 21.34 &  19.20 & 21.82 & \textbf{22.52} & 22.80  \\
        \vspace{0.2ex}
        & 0.558 & 0.523 & 0.524 & 0.537 & 0.552 & 0.441 & 0.622 & \textbf{0.622} & 0.665 \\
        \hline
        \multirow{2}{*}[0.2em]{$\alpha = 40$} & 17.84 & 20.56 & 20.51 & 21.21 & 21.53 & 17.34 & 21.76 & \textbf{22.67} & 22.99 \\
         & 0.565 & 0.533 & 0.531 & 0.545 & 0.566 &  0.382 & 0.634 & \textbf{0.638} & 0.690 \\
        \hline
        Blind? & \cmark & \cmark & \cmark &  \cmark &\cmark & \xmark & \xmark & \cmark & \xmark\\
        Network? & \xmark & \cmark & \cmark & \cmark &\cmark & \cmark & \xmark & \cmark & \cmark\\
        Unsupervised? & \cmark & \xmark & \xmark & \xmark & \xmark & \cmark & \cmark & \cmark & \xmark \\
        \midrule[0.3pt]\bottomrule[1pt]

    \end{tabular}
    \vspace{1ex}
    \caption{\textbf{Performance on Levin et al. Dataset} \cite{levin2011understanding}: \textit{(Top)} Average PSNR in dB \textit{(Bottom)} SSIM. The last column, P4IP, is a non-blind deconvolution method, used as $F(\cdot)$ in the iterative blind deconvolution method and serves as an upper bound of the iterative scheme. The gray-colored columns represent the unsupervised methods which cannot be trained end-to-end.}
    \label{tab:psnr_ssim}
\end{table*}

\begin{table}[]
    \setlength\doublerulesep{0.5pt}
    \centering
    \begin{tabular}{l|ccc}
        \hline
        \toprule[1pt]
        \multirow{2}{*}[0.5em]{Photon lvl $\rightarrow$} & \multirow{2}{*}[0.0em]{$\alpha = 10$} & \multirow{2}{*}[0.0em]{$\alpha = 20$} &
        \multirow{2}{*}[0.0em]{$\alpha = 40$}\Tstrut \\
        Method $\downarrow$ & & & \\
        \midrule[0.3pt]
        DMPHN \cite{zhang2019deep} &  23.88 & 24.26 & 24.20 \\
        SRN \cite{tao2018scale} & 23.89 & 24.05 & 24.00 \\
        Deep-Deblur \cite{Nah_2017_CVPR} & 24.57 & 24.45 & 24.65 \\
        MPRNet \cite{zamir2021multi} & 25.50 & 25.93 & 25.97 \\
        Ours & \textbf{26.49} & \textbf{27.43} & \textbf{26.67}  \\
        \midrule[0.3pt]\bottomrule[1pt]
    \end{tabular}
    \vspace{1ex}
    \caption{\textbf{Performance on Real-Blur Dataset} \cite{realblur}:  Average PSNR in dB.}
    \label{tab:real_blur}
\end{table}

\subsection{Training $F(\cdot)$ and $G(\cdot)$}
In this subsection, we describe the training procedure for the non-blind solver $F(\cdot)$ and the denoiser $G(\cdot)$. These training processes, along with other experiments described in paper, are implemented in PyTorch 1.7.0, and use an NVIDIA Titan Xp GP102 GPU.

For the non-blind solver $F(\cdot)$, we use a similar process as that of \cite{sanghvi2021photon}. We use 2650 clean images from the Flickr2K dataset \cite{Flickr2K} and divide them into a 80:20 training and validation dataset. We generate 60 motion kernels described using the code provided in \cite{motion_blur}. To generate synthetically blurred-noisy image,  clean images from the Flickr2K training subset are blurred by the motion kernels followed by the Poisson noise corruption at photon level $\alpha$. The photon level $\alpha$ is uniformly sampled from $[1, 60]$. Using the pairs of noisy-blurred images, corresponding blur kernels and the clean images, we train PhD-Net ($K = 8$ iterations)  from \cite{sanghvi2021photon} using the $\ell_1$ loss.

For training the blur-noisy to blur-only denoiser $G(\cdot)$, we use a similar method. We use synthetically generated noisy-blurred images from the training process of non-blind solver $F(\cdot)$. However, there are a few key distinctions here. We use the input as noisy-blurred images and don't use the relevant blur kernel. The target for training is \emph{not the clean image but the blur-only image}. For the network architecture, we use the same configuration as that of PhD-Net, but change we fix the kernel input to be the identity operator. We use the $\ell_1$ loss function to train the denoiser.

\subsection{Quantitative Comparison}
We compare our proposed iterative scheme with five other blind deconvolution methods in this paper: a classical Two-Phase Estimation \cite{xu2010two}, Scale-Recurrent Network \cite{tao2018scale}, Deep-Deblur \cite{Nah_2017_CVPR}, Deep-Hierarchical Multi-Patch Network \cite{zhang2019deep}, and MPR-Net  \cite{zamir2021multi}. When the ground-truth blur kernel is available, we also include the following non-blind deconvolution algorithms in the comparison: PURE-LET \cite{purelet}, Poisson PnP \cite{rond2016poisson}. Since Two-Phase estimation \cite{xu2010two} is an unsupervised method designed for noiseless images, we use the denoiser output $G(\vy)$ as the input to the algorithm instead of Poisson corrupted image $\vy$.

For the supervised end-to-end trainable methods i.e., \cite{tao2018scale,Nah_2017_CVPR,zhang2019deep,zamir2021multi}, we retrain the networks for deconvolving Poissonian images in the same manner as training $F(\cdot)$ and $G(\cdot)$. Specifically, we use the artificially generated clean and noisy-blurred image pairs from Fickr2K dataset as described in the previous subsection, are retrain them with Adam optimizer of learning rate $10^{-4}$, $\ell_1$ loss functions until training converges.

We remark that the training procedure outlined above is a fair setting for all the methods we consider in this paper. The reason is that supervised methods such as \cite{zamir2021multi,Nah_2017_CVPR,zhang2019deep,tao2018scale} have a an unfair advantage over our scheme because they can see the ground truth. The ground truths allow them to handle blurs beyond a simple spatially invariant blur we assume (which is also assumed by all other unsupervised blind deconvolution methods). Since in this paper we are focused on a spatially invariant blur, we argue that it is more useful to compare all methods using the same forward image formation model. Future work could address how the scheme can be modified to take into account more complex forward models such as spatially varying blur and blur due to object motion.

\textbf{Levin Dataset \cite{levin2011understanding}}: Once the supervised methods are retrained using Poisson data, we compare the different schemes on the Levin et al. dataset \cite{levin2011understanding} which consists of 4 images, each blurred by 8 different motion kernels. The Poisson noise is artificially added at different light levels $\alpha = 10, 20, 40$. The results of the comparison are summarized in Table \ref{tab:psnr_ssim}. Note that in Table \ref{tab:psnr_ssim}, we also compare the blind deconvolution methods with the non-blind deconvolution network P4IP \cite{sanghvi2021photon}. While this network is used iteratively in our scheme as $F(\cdot)$, we provide its performance using the ground truth blur kernel, to serve as an upper bound to the performance of blind deconvolution algorithms.

\textbf{Real-Blur Dataset \cite{realblur}}: In addition to the Levin dataset, we perform a quantitative evaluation on the Real-Blur dataset \cite{realblur} which contains pairs of sharp and blurred images for 232 different scenes with ~20 image pairs per scene. These pair of images are captured  using a specially designed image acquisition system that contains a beam splitter and two cameras - one for ground-truth operating at $1/80$s exposure time and the other at $1/2$s for the blurred images.

For our experiment, we evaluate different blind-deconvolution methods on a randomly chosen $256 \times 256$ patch from a  ground-truth, blurred pair from 50 different scenes in  \textit{RealBlur-J} subset. We degrade the blurred images with Poisson noise at photon level $\alpha = 10, 20, \text{ and } 40$ and the resulting PSNR values are shown in Table \ref{tab:real_blur}. Since the blur kernel is not available for this dataset, unlike Levin, the non-blind deconvolution methods i.e. Poisson-PnP \cite{rond2016poisson}, PURE-LET \cite{purelet}, P4IP \cite{sanghvi2021photon} cannot be evaluated on this dataset.

\begin{figure*}[!]
\centering
\begin{tabular}{ccccc}
 \multirow{2}[2]{*}[16.5mm]{\hspace{-2.0ex}\makecell{\includegraphics[width=0.36\linewidth]{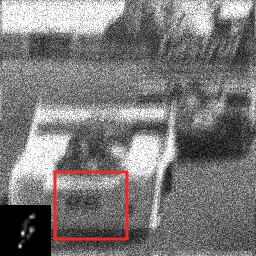} \\ Input \\ \small{16.35 / 0.280} }} &
\hspace{-1.5ex}\makecell{\includegraphics[width=0.15\linewidth]{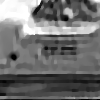} \\ Two-Phase \cite{xu2010two} \\ \small{17.82 / 0.640 } } &
\hspace{-1.5ex}\makecell{\includegraphics[width=0.15\linewidth]{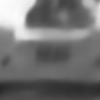} \\ SRN \cite{tao2018scale} \\ \small{18.89 / 0.459}} &
\hspace{-1.5ex}\makecell{\includegraphics[width=0.15\linewidth]{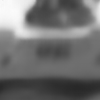} \\ DHMPN \cite{zhang2019deep} \\ \small{18.70 / 0.452}}
 & \hspace{-1.5ex}\makecell{\includegraphics[width=0.15\linewidth]{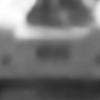} \\ Deep-Deblur \cite{Nah_2017_CVPR} \\ \small{18.85 / 0.455} } \\
 & \hspace{-1.5ex}\makecell{\includegraphics[width=0.15\linewidth]{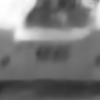} \\ MPR-Net \cite{zamir2021multi} \\ \small{19.11 / 0.477}} &
\hspace{-1.5ex}\makecell{\includegraphics[width=0.15\linewidth]{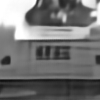} \\ \textbf{Ours} \\ \textbf{\small{19.86 / 0.599}} } &
\hspace{-1.5ex}\makecell{\includegraphics[width=0.15\linewidth]{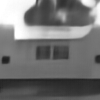} \\ Non-Blind \cite{sanghvi2021photon} \\ \small{21.93 / 0.621}} &
\hspace{-1.5ex}\makecell{\includegraphics[width=0.15\linewidth]{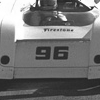} \\ Ground-Truth \\ \small{\phantom{-}} }
\vspace{2.0ex}
\end{tabular}
\begin{tabular}{ccccc}
 \multirow{2}[2]{*}[16.5mm]{\hspace{-2.0ex}\makecell{\includegraphics[width=0.36\linewidth]{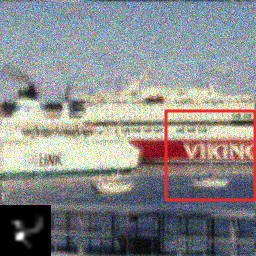} \\ Input \\ \small{15.72 / 0.278} }} &
\hspace{-1.5ex}\makecell{\includegraphics[width=0.15\linewidth]{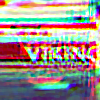} \\ Two-Phase \cite{xu2010two} \\ \small{12.99 / 0.599} } &
\hspace{-1.5ex}\makecell{\includegraphics[width=0.15\linewidth]{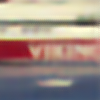} \\ SRN \cite{tao2018scale} \\ \small{17.84 / 0.486}}  &
\hspace{-1.5ex}\makecell{\includegraphics[width=0.15\linewidth]{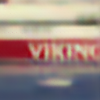} \\ DMPHN \cite{zhang2019deep} \\ \small{17.63 / 0.483}} & \hspace{-1.5ex}\makecell{\includegraphics[width=0.15\linewidth]{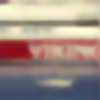} \\ Deep-Deblur \cite{Nah_2017_CVPR} \\ \small{17.69 / 0.475}} \\
 & \hspace{-1.5ex}\makecell{\includegraphics[width=0.15\linewidth]{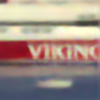} \\ MPR-Net \cite{zamir2021multi}\\ \small{18.48 / 0.551}} &
\hspace{-1.5ex}\makecell{\includegraphics[width=0.15\linewidth]{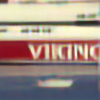} \\ \textbf{Ours} \\ \textbf{\small{19.76 / 0.639}} } &
\hspace{-1.5ex}\makecell{\includegraphics[width=0.15\linewidth]{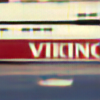} \\ Non-Blind \cite{sanghvi2021photon} \\ \small{20.74 / 0.652}} &
\hspace{-1.5ex}\makecell{\includegraphics[width=0.15\linewidth]{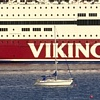} \\ Ground-Truth\\ \small{ \phantom{$\infty$ / 1.00} } }
\vspace{2.5ex}
\end{tabular}

\begin{tabular}{ccccc}
 \multirow{2}[2]{*}[16.5mm]{\hspace{-2.0ex}\makecell{\includegraphics[width=0.36\linewidth]{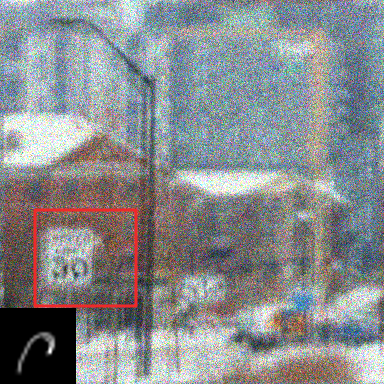} \\ Input \\ \small{14.73 / 0.191}}} &
\hspace{-1.5ex}\makecell{\includegraphics[width=0.15\linewidth]{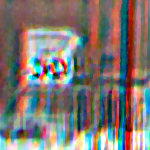} \\ Two-Phase \cite{xu2010two} \\ \small{14.73 / 0.191}} &
\hspace{-1.5ex}\makecell{\includegraphics[width=0.15\linewidth]{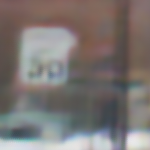} \\ SRN \cite{tao2018scale}\\\small{18.91 / 0.358}} &
\hspace{-1.5ex}\makecell{\includegraphics[width=0.15\linewidth]{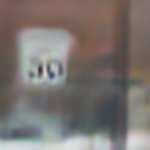} \\ DMPHN \cite{zhang2019deep}\\ \small{18.87 / 0.356}}
 & \hspace{-1.5ex}\makecell{\includegraphics[width=0.15\linewidth]{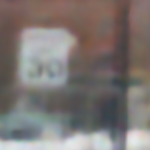} \\ Deep-Deblur \cite{Nah_2017_CVPR} \\ \small{19.11 / 0.365}} \\
 & \hspace{-1.5ex}\makecell{\includegraphics[width=0.15\linewidth]{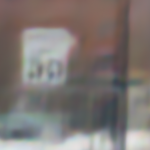} \\ MPR-Net \cite{zamir2021multi}\\\small{19.00 / 0.358}} &
\hspace{-1.5ex}\makecell{\includegraphics[width=0.15\linewidth]{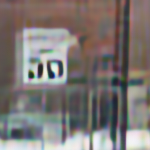} \\ \textbf{Ours} \\ \textbf{\small{20.03 / 0.459}} } &
\hspace{-1.5ex}\makecell{\includegraphics[width=0.15\linewidth]{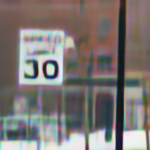} \\ Non-Blind \cite{sanghvi2021photon} \\ \small{21.99 / 0.549}} &
\hspace{-1.5ex}\makecell{\includegraphics[width=0.15\linewidth]{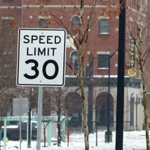} \\ Ground-Truth \\ \small{\phantom{-}}}
\vspace{4.0ex}
\end{tabular}
 \caption{\textbf{Visual comparisons}. Qualitative results of the competing methods on synthetically blurred and Poisson corrupted images from the BSD100 dataset. The specific blur kernel used in each experiment is shown in the inset of the input. The photon level for  \texttt{car}, \texttt{viking}, \texttt{speed-limit} is $\alpha = 40, 40$ and $20$ respectively.}
\label{fig:qual_comparison}
\end{figure*}

\begin{figure*}[ht]
\centering
\begin{tabular}{cccc}
 \hspace{-2.0ex}\multirow{2}[2]{*}[17mm]{\makecell{\includegraphics[width=0.345\linewidth]{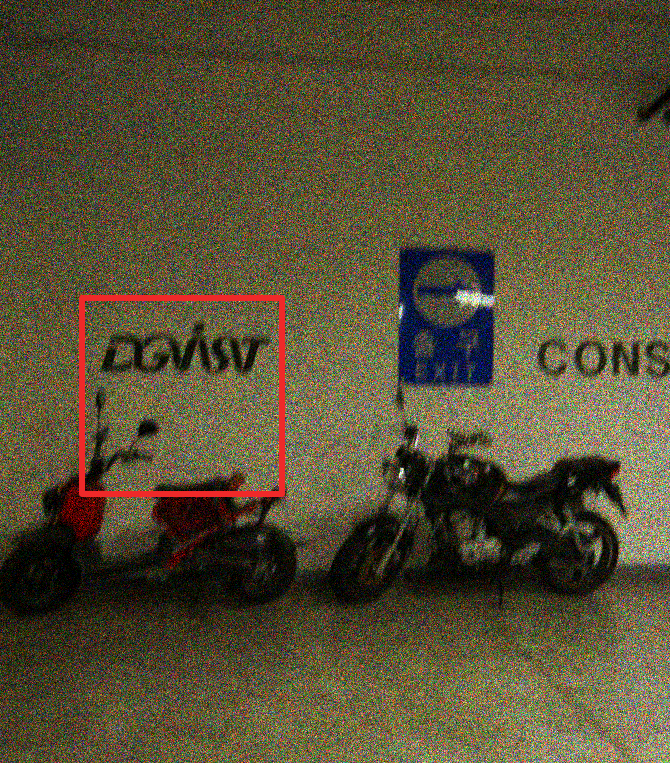} \\ Input }} &
\makecell{\includegraphics[width=0.18\linewidth]{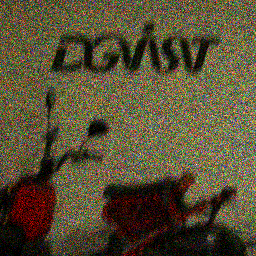} \\ Input } &
\hspace{-1.5ex}\makecell{\includegraphics[width=0.18\linewidth]{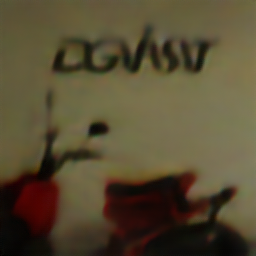} \\ SRN \cite{tao2018scale}} &
\hspace{-1.5ex}\makecell{\includegraphics[width=0.18\linewidth]{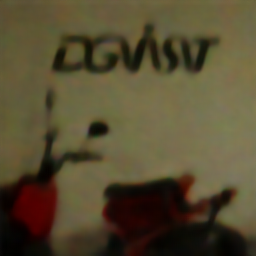} \\ Deep-Deblur \cite{Nah_2017_CVPR}}
\\
& \makecell{\includegraphics[width=0.18\linewidth]{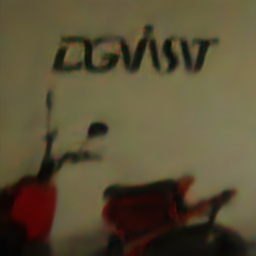} \\ MPR-Net \cite{zamir2021multi}} & \hspace{-1.5ex}\makecell{\includegraphics[width=0.18\linewidth]{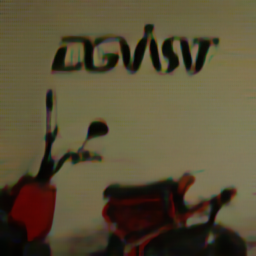} \\ \textbf{Ours}} &
\hspace{-1.5ex}\makecell{\includegraphics[width=0.18\linewidth]{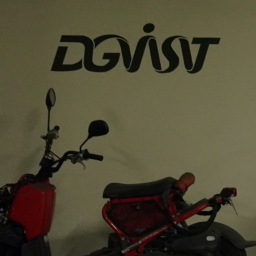} \\ Ground-Truth}
\end{tabular}
\vspace{4.0ex}

\begin{tabular}{cccc}
 \hspace{-2.0ex}\multirow{2}[2]{*}[17mm]{\makecell{\includegraphics[width=0.345\linewidth]{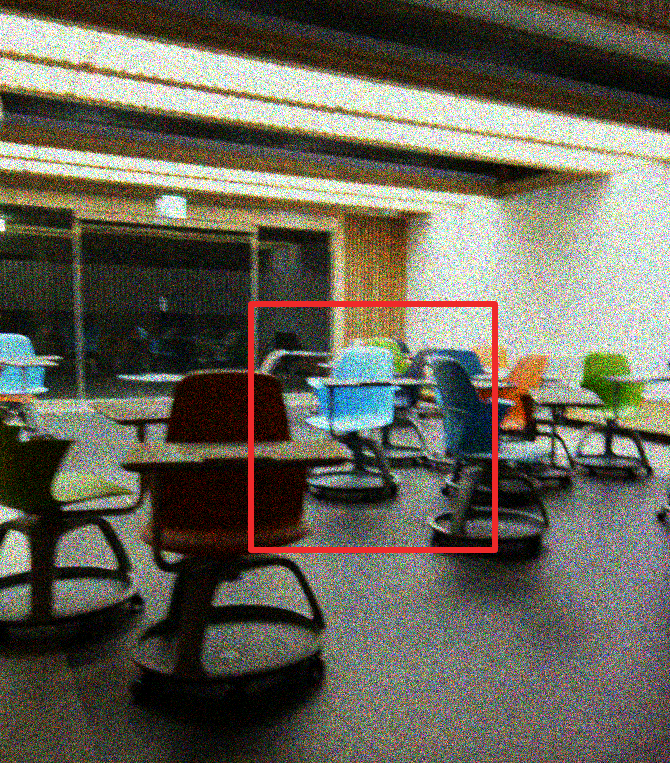} \\ Input }} &
\makecell{\includegraphics[width=0.18\linewidth]{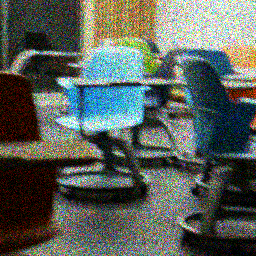} \\ Input } &
\hspace{-1.5ex}\makecell{\includegraphics[width=0.18\linewidth]{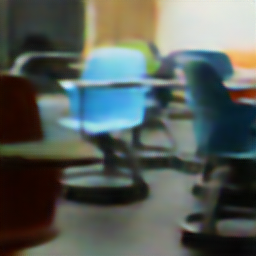} \\ SRN \cite{tao2018scale}} &
\hspace{-1.5ex}\makecell{\includegraphics[width=0.18\linewidth]{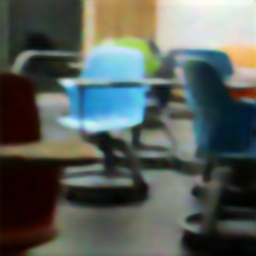} \\ Deep-Deblur \cite{Nah_2017_CVPR}}
\\
& \makecell{\includegraphics[width=0.18\linewidth]{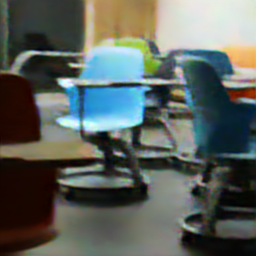} \\ MPR-Net \cite{zamir2021multi}} & \hspace{-1.5ex}\makecell{\includegraphics[width=0.18\linewidth]{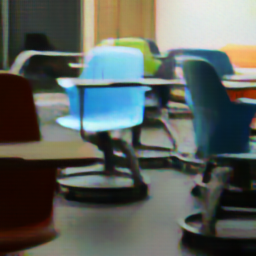} \\ \textbf{Ours}} &
\hspace{-1.5ex}\makecell{\includegraphics[width=0.18\linewidth]{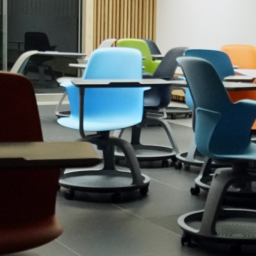} \\ Ground-Truth}
\end{tabular}

\vspace{4.0ex}

\caption{\textbf{Visual comparisons on Realistic Blur}. Qualitative results on realistically blurred images from the RealBlur dataset \cite{realblur}. Photon levels for the first and second images are $\alpha=20$ and $\alpha =10$ respectively. }
\label{fig:qual_comparison_realblur}
\end{figure*}

\begin{figure*}[]
    \centering
    \begin{tabular}{cccc}
        \includegraphics[width=0.245\linewidth]{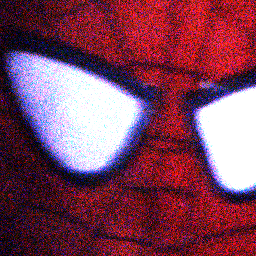} &
        \hspace{-1.5ex}\includegraphics[width=0.245\linewidth]{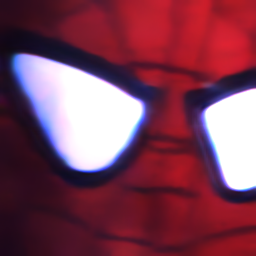} &
        \hspace{-1.5ex}\includegraphics[width=0.245\linewidth]{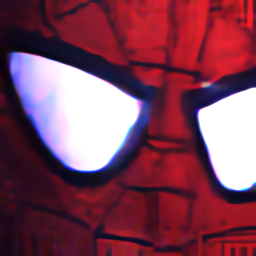} &
        \hspace{-1.5ex}\includegraphics[width=0.245\linewidth]{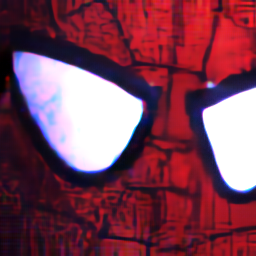} \\
        Blurred and Noisy & MPR-Net \cite{zamir2021multi} & \textbf{Ours} & Non-Blind \cite{sanghvi2021photon} \\
        \vspace{2.0ex}
        \includegraphics[width=0.245\linewidth]{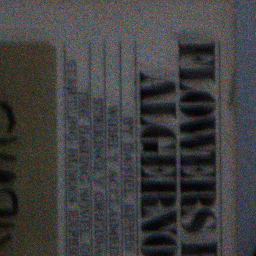} &
        \hspace{-1.5ex}\includegraphics[width=0.245\linewidth]{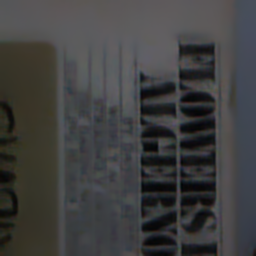} &
        \hspace{-1.5ex}\includegraphics[width=0.245\linewidth]{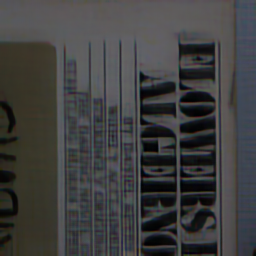} &
        \hspace{-1.5ex}\includegraphics[width=0.245\linewidth]{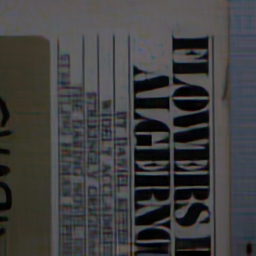} \\
        Blurred and Noisy & MPR-Net \cite{zamir2021multi} & \textbf{Ours} & Non-Blind \cite{sanghvi2021photon}
        \end{tabular}
    \caption{\textbf{Real world data} from \cite{sanghvi2021photon}. Note that the the blur kernel is unknown to the reconstruction methods except the non-blind solver in the last column.}
    \label{fig:real_images}
\end{figure*}

\subsection{Qualitative Comparison}
A qualitative comparison of grayscale and color reconstructions using different schemes is provided in Figure \ref{fig:qual_comparison}. We also provide the corresponding result using the non-blind solver from \cite{sanghvi2021photon} as a reference since this method assumes the ground truth kernel to be known. For reconstructing color images, we first describe how to simulate synthetic blur and noise and then how to modify the scheme from deconvolving grayscale to color images.

To simulate blur and Poisson noise in color images, we convert the RGB image to the corresponding Bayer pattern image - which itself can be viewed as 4 channels, namely \textit{R, G1, G2, B}, interleaved with each other. To simulate photon-limited blur, we blur each channel (R, G1, G2, and B) individually and add Poisson noise at the same photon level.

For our deconvolution scheme, we pick a single channel, say \textit{R}, and apply the kernel estimation process described in Section 3 to it. Using the kernel estimate obtained, we deconvolve each of the four channels with the non-blind solver $F(\vy, \vk)$. The deconvolved channels are combined into an RGB image using an off-the-shelf demosaicing method. For other schemes such as \cite{xu2010two,tao2018scale,Nah_2017_CVPR,zamir2021multi}, we perform deconvolution for each channel followed by demosaicing.

\subsection{Qualitative Evaluation on Real World Data}
We also demonstrate that our iterative scheme can reconstruct real-world blurred and noisy images. First we demonstrate our qualitative reconstructions on examples from the \textit{RealBlur} dataset in Figure \ref{fig:qual_comparison_realblur}. The blurred images in the dataset are degraded by Poisson shot noise at photon level $\alpha = 20$ and $\alpha = 10$.

Next, we demonstrate the iterative scheme proposed in this paper on patches from the dataset provided in \cite{sanghvi2021photon}. This dataset consists of 30 images collected using a DSLR camera with blur in the images generated using handheld motion. Due to the photon-limited setting of these scenes, the images are naturally corrupted by the photon shot noise, which follows a Poisson distribution. Hence, unlike the \textit{RealBlur} dataset, this contains both realistic blur and photon-shot noise.  While the blur kernel is also estimated using a point source, it is not used in the reconstruction scheme in this paper. For further details about the dataset, we refer the reader to \cite{sanghvi2021photon}. The reconstructions using the dataset are reported in Figure \ref{fig:real_images} along with the reconstruction using MPR-Net \cite{zamir2021multi} and the non-blind solver $F(\vy, \vk)$ which uses the ground truth kernel.

Note that for images with realistic blur, circular boundary conditions for convolution lead to artifacts and symmetric boundary conditions are more accurate to model real-world blur. As a result, the input to $F(\cdot)$ is padded symmetrically on both sides before the iterative scheme starts. The relevant center portion of the image is cropped out of the output $F(\vy, \vh)$.

\subsection{Kernel Estimation}

We quantitatively compare the estimated kernel. In Table \ref{tab:kernel_estimate}, we show the mean absolute error for each of the eight kernels in the Levin dataset \cite{levin2006blind}. The mean absolute error (MAE) is defined as follows:
\begin{align}
    \text{MAE} \bydef \|\widehat{\vh} - \vh\|_1/M
\end{align}
where $\widehat{\vh}$, $\vh$ are the estimated and the ground-truth kernels respectively, and $M$ represents the number of pixels of the kernel. We compare the estimated kernel with that of the Two-Phase estimation \cite{xu2010two} which takes $G(\vy)$ as the input and gives the estimated kernel. We observe a consistently better kernel estimate using our proposed method.

\begin{table*}[!]
\setlength\tabcolsep{1.5pt}
    \centering
    \setlength\doublerulesep{0.5pt}
    \begin{tabular}{c|cccccccccccccccc}
        \hline
        \toprule[1pt]
    \makecell{Kernel $\rightarrow$\\ Photon Level $\downarrow$ }&
    \multicolumn{2}{c}{\includegraphics[width=0.1\linewidth]{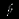}}&
    \multicolumn{2}{c}{\includegraphics[width=0.1\linewidth]{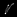}}&
    \multicolumn{2}{c}{\includegraphics[width=0.1\linewidth]{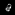}}&
    \multicolumn{2}{c}{\includegraphics[width=0.1\linewidth]{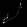}}&
    \multicolumn{2}{c}{\includegraphics[width=0.1\linewidth]{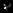}}&
    \multicolumn{2}{c}{\includegraphics[width=0.1\linewidth]{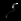}}&
    \multicolumn{2}{c}{\includegraphics[width=0.1\linewidth]{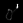}}&
    \multicolumn{2}{c}{\includegraphics[width=0.1\linewidth]{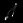}}\\
    & \cite{xu2010two} & Ours & \cite{xu2010two} & Ours & \cite{xu2010two} & Ours & \cite{xu2010two} & Ours &
    \cite{xu2010two} & Ours & \cite{xu2010two} & Ours & \cite{xu2010two} & Ours & \cite{xu2010two} & Ours\\
    \hline
    $\alpha = 10$ & 1.40 & \textbf{1.08} & 1.36 & \textbf{1.11} & 1.03 & \textbf{0.84} & 1.53 & \textbf{1.47} &
    1.13 & \textbf{0.92} & 1.38 & \textbf{1.16} & 1.47 & \textbf{1.27} & 1.46 & \textbf{1.29} \\
    $\alpha = 20$ & 1.37 & \textbf{1.15} & 1.28 & \textbf{1.12} & 0.91 & \textbf{0.86} & 1.51 & \textbf{1.40} &
    1.07 & \textbf{0.99} & 1.34 & \textbf{1.09} & 1.45 & \textbf{1.28} & 1.47 & \textbf{1.31} \\
    $\alpha = 40$ & 1.25 & \textbf{1.27} & 1.33 & \textbf{1.26 }& 0.94 & \textbf{0.88} & 1.52 & \textbf{1.42} &
    1.08 & \textbf{1.01} & 1.26 & \textbf{1.12} & 1.42 & \textbf{1.26} & 1.42 & \textbf{1.33}\\
    \midrule[0.3pt]\bottomrule[1pt]
    \end{tabular}
    \caption{Mean absolute error (MAE) on kernels estimated on the Levin dataset.}
    \label{tab:kernel_estimate}
\end{table*}

\section{Ablation Study}
\subsection{Non-Blind Solver}
Since a photon-limited non-blind deconvolution solver is at the core of the proposed scheme, we evaluate the performance of the scheme using different non-blind solvers. For this scheme, the non-blind solver $F(\cdot)$ can be any differentiable solver which takes as input a noisy-blurred image and a kernel input. Therefore, we compare the performance of the method on 3 different non-blind solvers in Table \ref{tab:nb_solver_ablation} - P4IP \cite{sanghvi2021photon}, DWDN \cite{dong2020deep}, and RGDN \cite{gong2020_LearningDeep}. We use the pre-trained P4IP model for the experiment and the latter non-blind solvers are retrained using Poisson data.

The results for this comparison are provided in Table \ref{tab:nb_solver_ablation}. We can see that the performance on the non-blind and blind deconvolution task are correlated to each other. Since P4IP has the best performance on non-blind deconvolution task, it outperforms other methods on the blind-deconvolution problem as well. However, while DWDN outperforms RGDN in the non-blind task, it only outperforms RGDN in blind deconvolution at photon level $\alpha = 40$.
\begin{table}[]
    \centering
    \setlength\doublerulesep{0.5pt}
    \begin{tabular}{cc|  c c c}
    \toprule[1pt]
      \hspace{-2.0ex}
    \multirow{2}{*}[0.5em]{Photon Level $\rightarrow$} & Blind / & \multirow{2}{*}[0.0em]{$\alpha = 10$} & \multirow{2}{*}[0.0em]{$\alpha = 20$} & \multirow{2}{*}[0.0em]{$\alpha = 40$}  \Tstrut\\
     Solver $\downarrow$ & Non-Blind & &  &  \Bstrut \\
     \midrule[0.3pt]
    \multirow{2}{*}[0.0em]{P4IP \cite{sanghvi2021photon}} &  Non-Blind & 24.70 & 25.89 & 27.10  \Tstrut\\
          & Blind & 22.26 & 22.92 & 22.99 \Bstrut \\
    \multirow{2}{*}[0.0em]{DWDN \cite{dong2020deep}} &  Non-Blind & 23.93 & 24.87 & 25.75  \Tstrut\\
          & Blind & 19.03 & 19.87 & 23.20 \Bstrut \\
    \multirow{2}{*}[0.0em]{RGDN \cite{gong2020_LearningDeep}} &  Non-Blind & 22.41 & 22.97 & 23.44  \Tstrut\\
          & Blind & 21.95 & 22.46 & 22.68 \Bstrut \\
          \midrule[0.3pt]
          \bottomrule[1pt]
    \end{tabular}
    \caption{\textbf{Ablation Study for Effect of Non-Blind Solver} The output PSNR (in dB) on Levin dataset for different non-blind solvers (retrained for Poisson noise) when plugged into the iterative blind deconvolution scheme.}
    \label{tab:nb_solver_ablation}
\end{table}

\subsection{Hyperparameters}
In addition to studying the effect of non-blind solvers, we also evaluate the effect of different hyperparameters on the performance of the scheme in Table \ref{tab:ablation}. The experiments in this subsection can be divided into 3 different categories. (i) We evaluate the effect of the denoiser $G(\cdot)$ and $\ell_1$ kernel prior on the performance of the iterative method. (ii) We then evaluate how the performance varies with different strength of the $\ell_1$ kernel prior by varying the parameter $\gamma$. (iii) We also vary the maximum number of iterations of the iterative scheme.

From Table \ref{tab:ablation}, we observe that denoising the image $G(\vy)$ is an important feature of the scheme and improves the performance by approximately $4$dB when used as a target for the $\ell_2$ loss. Compared to denoiser $G(\cdot)$, the $\ell_1$ kernel prior doesn't have as significant improvement on the scheme in lower photon levels i.e. $\alpha = 10, 20$. However, at $\alpha = 40$, it adds another $1.5$dB to the blind-deconvolution performance.  From the third experiment, we can infer that performance of the scheme slightly degrades after $I = 20$ iterations.
\begin{table*}
\centering
\setlength\doublerulesep{0.5pt}

    \centering
    \begin{tabular}{l|cc|cc|cc|cc}
        \toprule[1pt]
        \multirow{2}{*}[0.5em]{Photon Level $\rightarrow$} &  \multicolumn{2}{c}{$\alpha = 10$}  & \multicolumn{2}{c}{$\alpha = 20$} &  \multicolumn{2}{c}{$\alpha = 40$} & \multicolumn{2}{c}{$\alpha = 60$}    \Tstrut\\
        Ablation Study $\downarrow$ & PSNR & SSIM &
        PSNR & SSIM & PSNR & SSIM & PSNR & SSIM \Bstrut\\
        \midrule[0.1pt]
        w/o $\ell_1$ kernel prior, w/o denoiser $G(\cdot)$ & 17.32 & 0.375 &  18.23 & 0.406 & 18.83 & 0.447 & 19.24 & 0.470 \\
        w/o denoiser $G(\cdot)$ & 18.41 & 0.402 & 19.75 & 0.450 & 20.90 & 0.508 & 21.92 & 0.558 \\
        w/o $\ell_1$ kernel prior & 22.36 & 0.614 & 23.03 & 0.645 & 21.40 & 0.447 & 23.05 & 0.657 \\
        \midrule[0.1pt]
        with $\ell_1$ kernel prior, $\gamma = 10^{-3}$ & 22.38 & 0.590 & 22.78 & 0.613 & 22.93 & 0.630 & 23.03 & 0.643 \\
        with $\ell_1$ kernel prior, $\gamma = 10^{-4}$ & 22.50 & 0.600 & 22.95 & 0.626 & 23.02 & 0.644 & 23.04 & 0.654  \\
        with $\ell_1$ kernel prior, $\gamma = 10^{-5}$ & 22.44 & 0.606 & 22.98 & 0.631 & 22.64 & 0.660 & 23.04 & 0.656 \\
        with $\ell_1$ kernel prior, $\gamma = 10^{-6}$ & 22.29 & 0.610 & 22.96 & 0.639 & 21.59 & 0.642 & 23.06 & 0.658\\
        \midrule[0.1pt]
        Initialization only, no iterations & 18.20 & 0.466 & 20.07 & 0.563 & 19.64 & 0.561 & 22.73 & 0.657  \\
        maximum iterations $I = 10$ & 22.38 & 0.613 & 22.99 & 0.642 & 22.89 & 0.663 & 23.04 & 0.654 \\
        maximum iterations $I = 20$ & 22.56 & 0.608 & 23.03 & 0.636 & 23.08 & 0.660 & 23.04 & 0.654 \\
        maximum iterations $I = 100$ & 22.29 & 0.597 & 22.98 & 0.630 & 23.03 & 0.644 & 23.04 & 0.654 \\
        \bottomrule[1pt]

    \end{tabular}
    \vspace{1ex}
    \caption{\textbf{Ablation Study} we perform ablation studies to better understand the impact of different hyperparameters on the iterative scheme. First, we test the performance of the scheme without using the denoiser $G(y)$ as a target and without using the $\ell_1$ prior. We study the effect of varying the strength of the $\ell_1$ prior i.e. hyperparameter $\gamma$ and changing the number of iteration of the scheme on the performance. }
    \label{tab:ablation}
\end{table*}

\section{Conclusion}
This paper presented a new iterative scheme for photon-limited blind deconvolution. The success of the method depends on the following characteristics of the proposed scheme which were absent in previous methods: (1) A differentiable and powerful non-blind solver as the backbone. While classical methods only use hand-crafted image priors, we use the latest deep learning based non-blind Poisson solver to estimate the latent image. This new solver is differentiable and thus allows back-propagation. (2) Alternating minimization is known to have difficulties when noise is strong. We re-formulate the problem as an equilibrium statement by absorbing the non-blind Poisson solver into the kernel estimation process. Experimental results confirm the effectiveness of the method.

Two aspects of the algorithm can be improved: (1) The prior of the kernel can be better chosen, as the $\ell_1$-norm prior is suitable only for a subset of motion blurs. A more powerful learning-based prior can be considered. (2) The computational cost needs to be reduced as backpropagation through a large non-blind solver requires a lot of memory. Some approximation schemes could be useful to speed up the computation.

\bibliographystyle{IEEEtran}
\bibliography{egbib}
\end{document}

\author{Yash~Sanghvi,~\IEEEmembership{Student~Member,~IEEE}, Abhiram~Gnanasambandam,~\IEEEmembership{Student~Member,~IEEE}, Zhiyuan~Mao,~\IEEEmembership{Student~Member,~IEEE}, and~Stanley~H.~Chan,~\IEEEmembership{Senior~Member,~IEEE}%
\thanks{Y.~Sanghvi, Z.~Mao and S.~Chan are with the School of Electrical and Computer
Engineering, Purdue University, West Lafayette, IN 47907, USA. The work of A.~Gnanasambandam was completed when he was a graduate student at Purdue University.  Email: {
\{ysanghvi, mao114, stanchan\}}@purdue.edu,  abhiram.g94@gmail.com } 
}

\title{Photon-Limited Blind Deconvolution using Iterative Kernel Estimation - Supplementary} % Replace with your title

\maketitle
\section{Real-World Dataset Utilization}
To reconstruct real blurred and noisy images from the dataset provided in \cite{realblur, sanghvi2021photon}, our scheme needs to modified in a minor way to account for realistic blur. 

We need to take into account that realistic blur varies from the synthetic blur in terms of boundary conditions for convolution. Assuming a circular boundary condition for convolution has computational advantage as it allows convolution to be written as an FFT operation which greatly speeds the operation. However, for realistic blur, symmetric boundary conditions are a much better approximation. To adjust our iterative scheme for symmetric boundary condition, whenever we use the non-blind solver $F(\vy, \vh)$, we pad the input image $\vy$ symmetrically before passing it as input to the non-blind solver. Then the relevant center portion of the output is cropped out and then used in the $\ell_2$ loss.

Secondly, with the real-noise and blur dataset in \cite{sanghvi2021photon}, the photon level $\alpha$ is not known and we use the heuristic provided by authors of the dataset to estimate it from the raw image itself which can be given as follows:
\begin{align}
    \hat{\alpha} = \frac{\sum\limits_{i=1}^N \vy_{i}}{\beta N}
\end{align}
i.e. the photon level $\alpha$ is estimated to be equal to the average photon-per-pixels divided by a constant factor $\beta = 0.33$.

\section{PyTorch Example Code}
To demonstrate with further clarity how we use autograd tools provided by PyTorch, we provide a code snippet of a single iteration of the iterative scheme proposed method. In this code snippet, we show how a "forward pass" is used to calculate the gradient term $\nabla_{\vh}\big\{L(\vh)\big\}\Big|_{\vh = \vh^k}$ from equation (13) in the main text. Note that this snippet is only for illustrative purposes and the full code for the scheme will be released as Github repository once the paper is accepted. 
\begin{figure}[H]
\begin{python}
"""
Code Snippet showing single iteration of scheme
yt: input image, ALPHA: photon level
h, h1: kernel estimates, 
"""
# Send to non-blind solver
x_out = p4ip(yt, h, ALPHA)
# Reconstruct the blurred image
y_rec = conv_fft_batch(h, x_out)
# MSE with denoised-only image as target
loss = torch.nn.MSELoss(yn_t, y_rec) 
# This step calculates the gradient of 
# h in variable h.grad 
loss.backward() 
# The following steps don't require 
# to be backpropagated through
with torch.no_grad():
    # Gradient descent step
    del_h = h.grad
    del_h += MU*(h-h1)
    h = h.sub_(STEP_SIZE*del_h)
    # Clip negative values to 0 and 
    # normalize kernel to 1 
    h = NORMALIZER(h); 
    h.requires_grad = True 
    # Shrinkage step
    h1 = shrinkage_torch(h, GAMMA/MU) 
\end{python}
\caption*{Code snippet demonstrating use of \texttt{autograd} tools of \texttt{PyTorch} for calculating gradient of $\vh$ w.r.t. loss function $\mathcal{L}(h)$}
\end{figure}

\begin{figure*}
    \hspace{-5.0ex}
    \begin{tabular}{cccc}
        \includegraphics[width=0.24\linewidth]{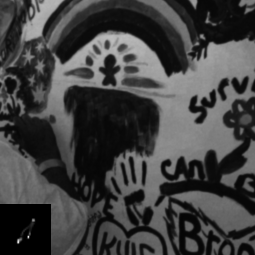} & 
        
        \includegraphics[width=0.24\linewidth]{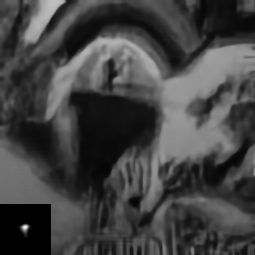} &
        
        \includegraphics[width=0.24\linewidth]{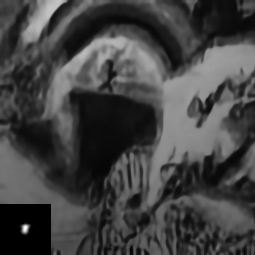} & 
        
        \includegraphics[width=0.24\linewidth]{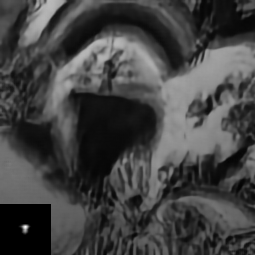} \\
        \includegraphics[width=0.24\linewidth]{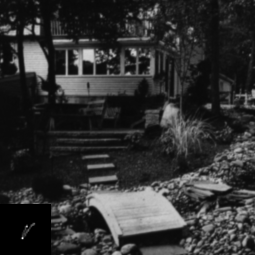} & 
        
        \includegraphics[width=0.24\linewidth]{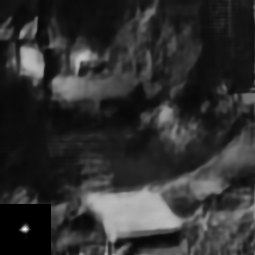} &
        
        \includegraphics[width=0.24\linewidth]{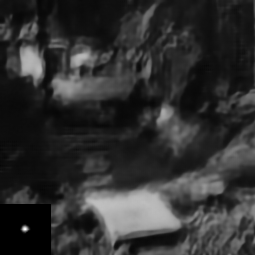} & 
        
        \includegraphics[width=0.24\linewidth]{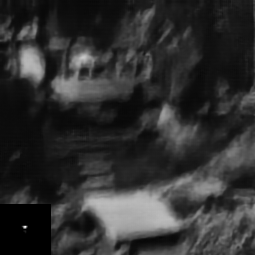} \\

        \includegraphics[width=0.24\linewidth]{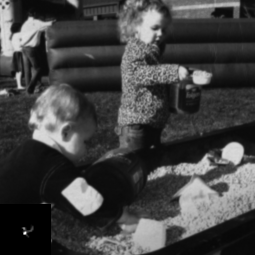} & 
        
        \includegraphics[width=0.24\linewidth]{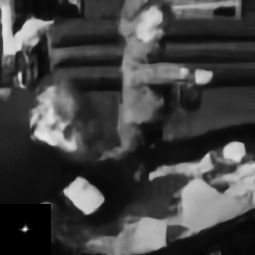} &
        
        \includegraphics[width=0.24\linewidth]{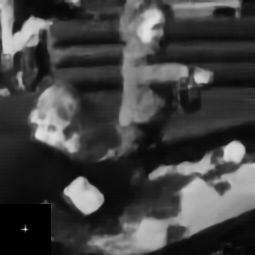} & 
        
        \includegraphics[width=0.24\linewidth]{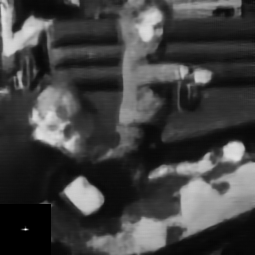} \\

        \includegraphics[width=0.24\linewidth]{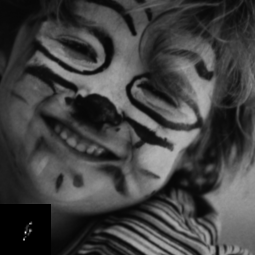} & 
        
        \includegraphics[width=0.24\linewidth]{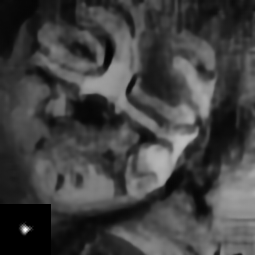} &
        
        \includegraphics[width=0.24\linewidth]{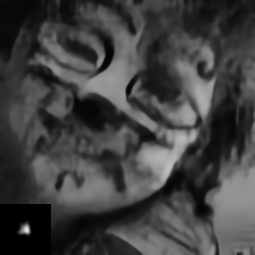} & 
        
        \includegraphics[width=0.24\linewidth]{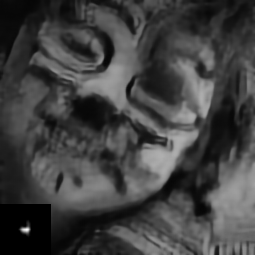} \\

        \makecell{True image, \\ True kernel in inset \\ Estimated images and kernels $\rightarrow$} & $\alpha = 10$ & $\alpha = 20$ & $\alpha = 40$ 
    \end{tabular}
    \caption{Example reconstructions from the Levin dataset along with true and estimated kernels in the inset.}
\end{figure*}

\section{Convergence Analysis}
\begin{figure*}[ht]
\centering
    \begin{subfigure}{0.17\linewidth}
    \centering
    \includegraphics[width=0.99\linewidth]{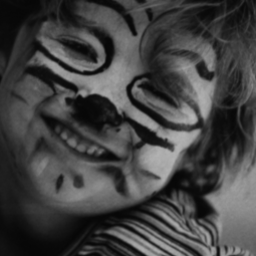} \\ \small{True image} \\
    \includegraphics[width=0.99\linewidth]{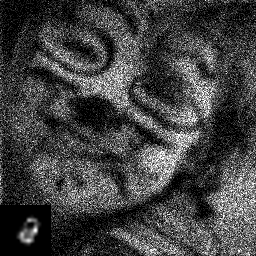} \\ \small{Blurred and noisy} 
    \end{subfigure}
    \begin{subfigure}{0.82\linewidth}
    \centering
        \includegraphics[trim={20 0 20 40},clip,width=0.99\linewidth]{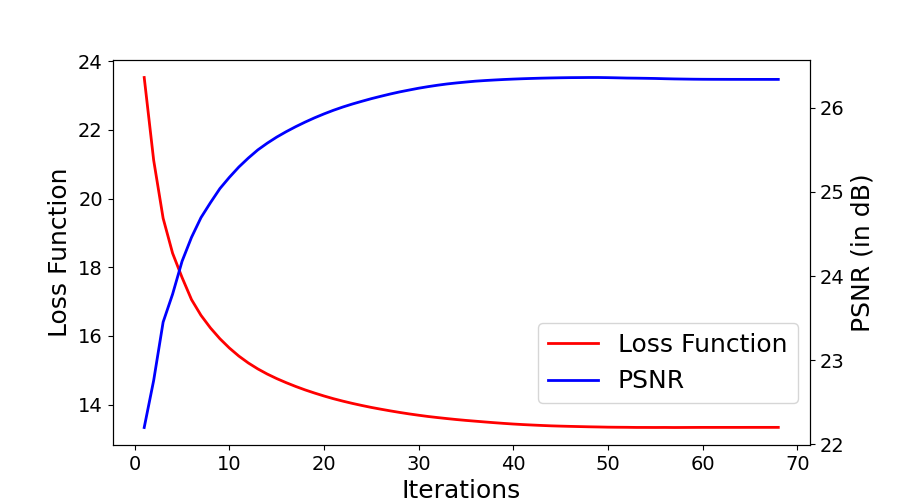} \\ 
        Loss function and PSNR across iterations
    \end{subfigure}
     \par\bigskip 
\centering
    \begin{tabular}{ccccc}
        \makecell[b]{ \small{Kernel and} \\ \small{Image $\rightarrow$} \\ \small{Cost Function} \\ \small{and PSNR $\downarrow$} \\  \\  } &
        \includegraphics[width=0.2\linewidth]{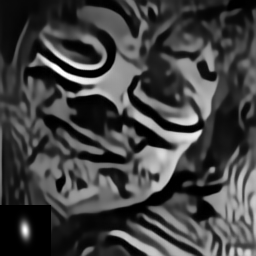} & 
        \includegraphics[width=0.2\linewidth]{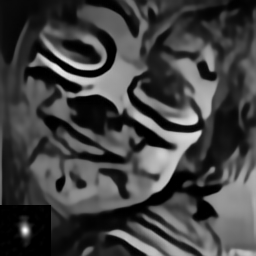} &
        \includegraphics[width=0.2\linewidth]{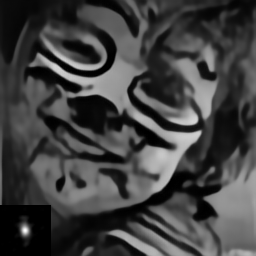} &
        \includegraphics[width=0.2\linewidth]{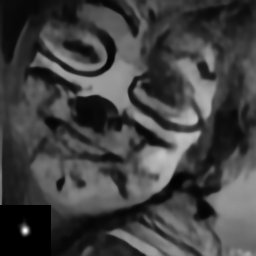}  \\
        Iteration $k$ & 0 & 5 & 10 & $\infty$ \\
        PSNR$^{k}$ & 22.20 & 24.46 & 25.29 & 26.33 \\
        $\mathcal{L}(\vh^{k})$ & $23.5 \times 10^{-5}$ & $17.0 \times 10^{-5}$ & $15.4 \times 10^{-5}$ & $13.3 \times 10^{-5}$ 
    \end{tabular}
\caption{Estimated kernel and corresponding image across iterations. $k = 0$, $k = \infty$ represent the kernel estimates from the initialization process and end of the iterative scheme respectively. }
\label{fig:ablation_study}
\end{figure*}

In Figure \ref{fig:ablation_study}, we provide a comparison of the kernel and corresponding image estimates over different iterations of the scheme for a single example, i.e., the initial kernel estimate (as obtained from the process described in Section 3.4), estimates at iterations $k = 10, 40$ and the final solution to which the scheme converges. We also report the value of loss function and PSNR for the corresponding iterations.

In the top half of Figure \ref{fig:ablation_study}, the loss function and PSNR are plotted across the iterations. The loss function decreases  monotonically while the PSNR increases upto iterations $k = 40$ followed by a small decline. This is also observed in Table V of the main document where setting the maximum number of iterations to $k = 20$ doesn't affect the performance by a lot.

\bibliographystyle{IEEEtran}
\bibliography{egbib}